\documentclass[a4paper,11pt]{article}
\pdfoutput=1 

\usepackage{jinstpub} 

\usepackage{caption}
\usepackage{subcaption}

\title{\boldmath Production and optical properties of liquid scintillator for the JSNS$^{2}$ experiment}

\author[a, 1]{J. S. Park,}\note{e-mail : jspark@post.kek.jp}
\author[d]{S. Y. Kim,}
\author[f]{C. Rott,}
\author[d]{D. H. Lee,}
\author[f]{D. Jung,}
\author[b]{F. Suekane,}
\author[a]{H. Furuta,}
\author[j]{H. I. Jang,}
\author[f]{H. K. Jeon,}
\author[f]{I. Yu,}
\author[h]{J. H. Choi,}
\author[i]{J. S. Jang,}
\author[e]{K. K. Joo,}
\author[g]{K. W. Ju,}
\author[h]{M. Pac,}
\author[e]{P. J. Gwak,}
\author[d]{S. B. Kim,}
\author[c]{S. Hasegawa,}
\author[f]{S. H. Jeon,}
\author[a]{T. Maruyama,}
\author[b]{R. Ujiie,}
\author[b,c]{Y. Hino,}
\author[e]{Y. S. Park}

\affiliation[a]{High Energy Accelerator Research Organization (KEK), Ibaraki, Japan}
\affiliation[b]{Research Center for Neutrino Science, Tohoku University, Sendai, Japan}
\affiliation[c]{Advanced Science Research Center, JAEA, Ibaraki, Japan}
\affiliation[d]{Department of Physics \& Astronomy, Seoul National University, Seoul, 08826, Korea}
\affiliation[e]{Institute for Universe \& Elementary Particles, Chonnam National University, Gwangju, 61186, Korea}
\affiliation[f]{Department of Physics, Sungkyunkwan University, Suwon, 16419, Korea}
\affiliation[g]{Korea Advanced Institute of Science and Technology, 34141, Korea}
\affiliation[h]{Institute for High Energy Physics, Dongshin University, Naju, 58245, Korea}
\affiliation[i]{Gwangju Institute of Science and Technology, Gwangju, 61005, Korea}
\affiliation[j]{Department of Fire Safety, Seoyeong University, Gwangju 61268, Korea}

\abstract{The JSNS$^{2}$ (J-PARC Sterile Neutrino Search at J-PARC Spallation Neutron Source) experiment will search for neutrino oscillations over a 24 m short baseline at J-PARC. The JSNS$^{2}$ inner detector will be filled with 17 tons of gadolinium-loaded liquid scintillator (LS) with an additional 31 tons of unloaded LS in the intermediate $\gamma$-catcher and outer veto volumes. JSNS$^{2}$ has chosen Linear Alkyl Benzene (LAB) as an organic solvent because of its chemical properties. The unloaded LS was produced at a refurbished facility, originally used for scintillator production by the RENO experiment. JSNS$^{2}$ plans to use ISO tanks for the storage and transportation of the LS. In this paper, we describe the LS production, and present measurements of its optical properties and long term stability. Our measurements show that storing the LS in ISO tanks does not result in degradation of its optical properties.}


\begin{document}
\maketitle
\flushbottom

\section{Introduction}
The JSNS$^{2}$ (J-PARC Sterile Neutrino Search at J-PARC Spallation Neutron Source) experiment aims to search for sterile neutrinos at a 24 m baseline with $\Delta$m$^{2}$ near $\sim$1 eV$^{2}$~\cite{cite:JSNS2_proposal}. JSNS$^{2}$ will use an intense source of muon antineutrinos from muon decay-at-rest; the muons are produced by the interactions of 3 GeV protons with a mercury target at the J-PARC Material and Life Science Experimental Facility. JSNS$^{2}$ will search for $\overline{\nu}_{\mu}$ $\rightarrow$ $\overline{\nu}_{e}$ oscillations, which can be detected via inverse beta decay, $\overline{\nu}_{e} + p \rightarrow e^{+} + n$. The detector consists of an inner target volume, an intermediate $\gamma$-catcher region, and an outer veto~\cite{cite:JSNS2_TDR, cite:JSNS2_Veto_detector}. The target region will be filled with 17 tons of gadolinium loaded liquid scintillator (LS), which will be donated by the Daya Bay experiment~\cite{cite:DayaBay_GdLS}. The optically separated $\gamma$-catcher and veto will be filled with 31 tons of LS. Linear alkyl benzene (LAB) was chosen as the organic solvent due to safety considerations and because of its high flash point, large attenuation length, and sufficient light yield. We produced the LS using a mass production facility from the RENO experiment. JSNS$^{2}$ plans to use ISO (International Organization for Standardization) tanks to store LS and transport it from the RENO site to Japan. The ISO tanks are made according to the ISO international standard, and the surface in contact with the LS is stainless steel (SUS316). Therefore, we do not expect any material compatibility issue.

\section{Mass production of liquid scintillator}
\subsection{Refurbishment of the RENO mass production facility}
\begin{figure}[h]
\begin{center}
\includegraphics[scale=0.5]{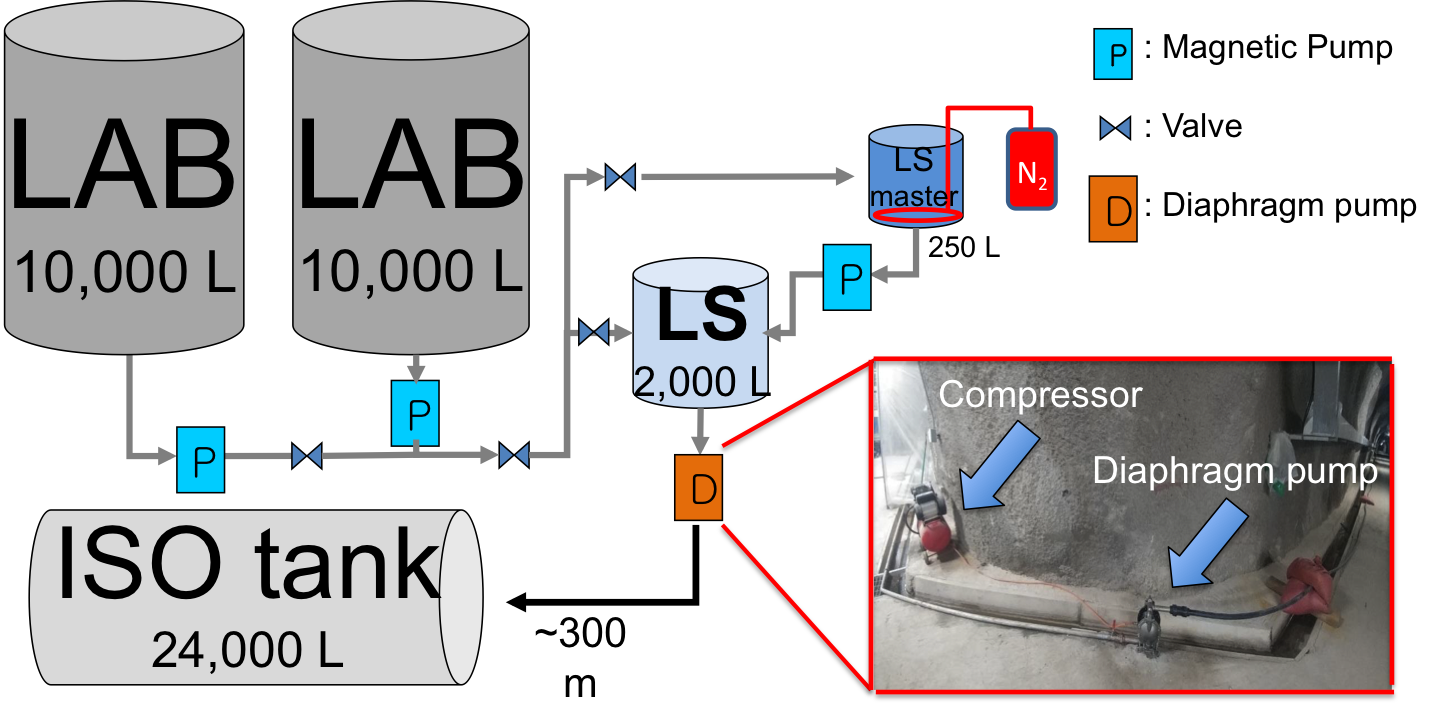}
\end{center}
\caption{\setlength{\baselineskip}{4mm}Conceptual drawing of the refurbished RENO LS mass production facility. LAB is stored inside of two 10,000 L stainless steel tanks and delivered to the acrylic production vessels by a magnetic pump. The finished LS is transferred outside via $\sim$300 m of 1-inch pipe, directly into an ISO tank.} 
\label{fig:refurbish}
\end{figure}

The RENO LS mass production facility consists of two of 10,000 L stainless steel LAB storage tanks, two acrylic production vessels (2000 L and 250 L), pipelines, stirring motors, impellers, magnetic pumps, and a nitrogen purging system. Two 24 kL ISO tanks were used to transfer the LS to Japan. The LS production facility was unused for 7 years and the pipelines were optimized for RENO, so the facility required cleaning and design modifications before it could be employed to make the JSNS$^{2}$ LS. A 3-step cleaning procedure was used for each tank and vessel. The initial cleaning was done with normal soap to remove organic contaminants and dust. Next, we used Alconox, which is a special powder detergent to remove deposits, chemicals, and radioactive contamination~\cite{cite:alconox}. Finally, everything was cleaned several times with high-pressure water to remove any remaining soap or Alconox. To enable safe operation of the facility, we purchased new stirring motors, impellers, and magnetic pumps. Due to size limitations, the ISO tanks were placed outside of the RENO entrance tunnel, and as a result, the pipelines were extended $\sim$300 m to reach the tanks. Additionally, we installed a strong diaphragm pump with an air compressor to transfer the LS from the production facility to the ISO tank. Figure~\ref{fig:refurbish} shows conceptual drawings of the refurbished facility.

\subsection{Mass production of LS}
The LS components and mass production method are similar to those used in RENO~\cite{cite:RENO_method}. We chose 3 g/L of PPO (2,5-diphenyloxazole) as the primary fluor and 30 mg/L of bis-MSB (1,4-bis(2-methylstyryl) benzene) as the secondary wavelength shifter. We produced 1800 L of LS in each batch. For simplicity, the LS is produced beginning with a master batch with 10 times the concentration of bis-MSB and PPO. To produce the master, we added $\sim$100 L of LAB to the 250 L acrylic production vessel, turned on a stirring motor, and added 54 g of bis-MSB. PPO is not added at this stage because the solubility of bis-MSB in LAB is lower than PPO. It took approximately 1 hour for the bis-MSB to dissolve into the LAB. Next, we added $\sim$80 L of LAB and 5.4 kg of PPO and stirred 4 additional hours. After stirring the mixture, $\sim$30 psig of nitrogen gas were injected to purge master solution. Finally, the master solution was transferred to the 2000 L acrylic vessel, and an additional 1800 L of LAB was added using a magnetic pump. The combined mixture was stirred for an additional 5 minutes in the 2000 L acrylic vessel to ensure uniform mixing of LS master and LAB. We took a 100 mL sample of each batch to measure the optical properties before pumping the LS into the ISO tank. A total of 38000 L of LS was produced in 21 batches including a small amount of extra scintillator (a few percent of the total). 

\section{Optical properties of LS}
\subsection{Water content}
Measuring the LS water content is an indirect method to assess the purging efficiency as shown by RENO~\cite{cite:RENO_method}. Purging the LS with nitrogen serves two main purposes: removing oxygen which may quench the scintillation and removing tiny bubbles which form in the scintillator. These tiny bubbles appear after several hours of powerful stirring during LS production and make the LS appear cloudy. It may be the case that these bubbles originate from small amounts of water vapor or oil and would dissipate over time. However, we decided to remove them forcibly by purging the LS. We measured the water content of each sample at Chonnam National University using a Mettler-toledo CH/C20 Coulometric Karl-Fischer Titrator, which can precisely measure the water content of the LS in the range 1 - 50000 ppm. The measured water content of each sample was 50 - 60 ppm, which is a similar to the RENO experiment meaning the amount of nitrogen purging was sufficient. 

\subsection{Relative Transmittance}
\begin{figure}[h]
\begin{center}
\includegraphics[scale=0.5]{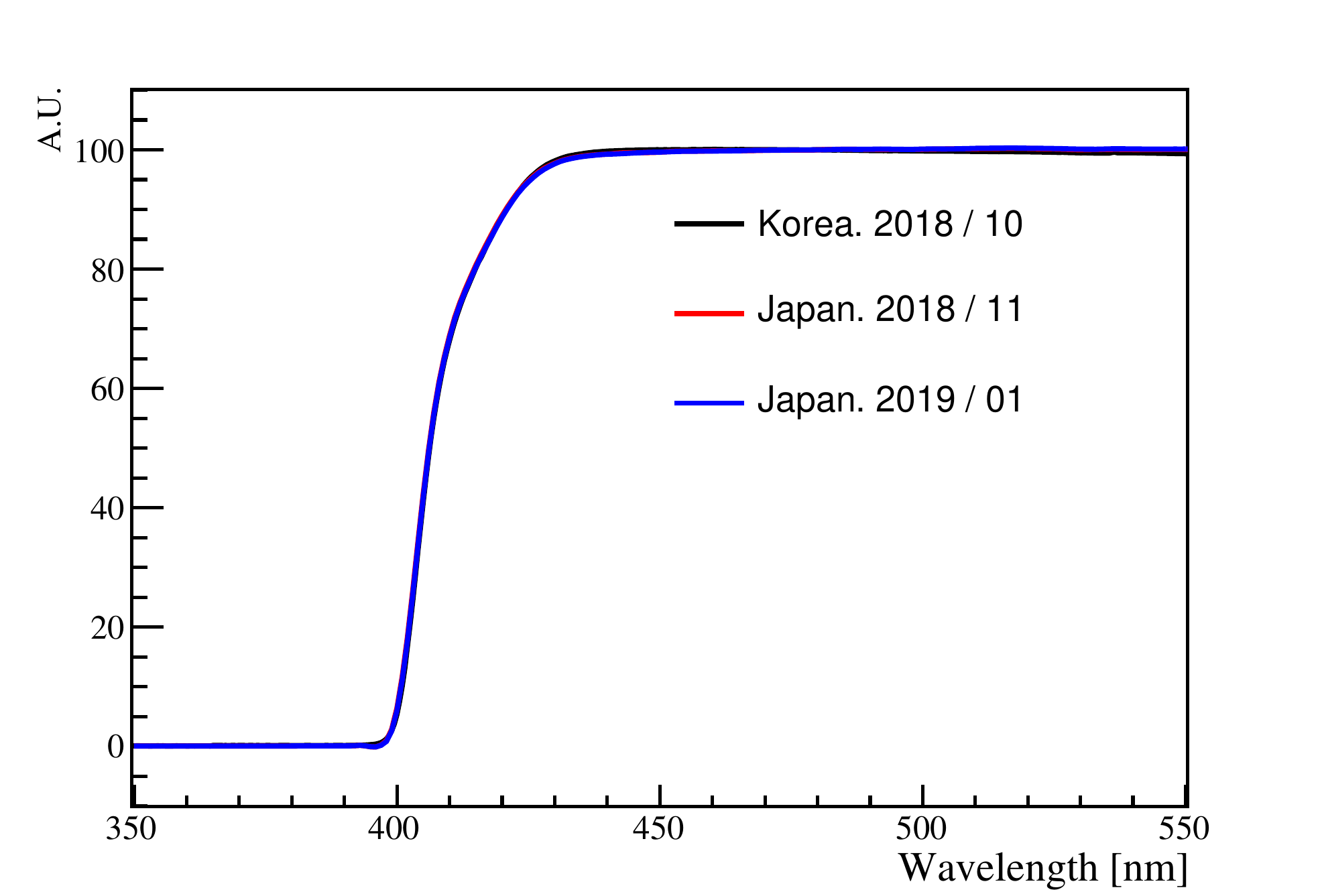}
\end{center}
\caption{\setlength{\baselineskip}{4mm}Transmittance measurement results in Korea a few days after production  (black line), and in Japan (red and blue lines). The LS was sampled twice in Japan with 2 months between the samples. The data from each measurement was normalized to 100 at 440 nm to remove the baseline fluctuations of the spectrometers. Each measurement shows the same structure, which means no significant degradation occurred during transit or during two months of storage inside of the ISO tanks.} 
\label{fig:transmittance}
\end{figure}

The transmittance of each LS sample was also measured at Chonnam National University. The measurements were done a few days after production with a 10 cm cuvette and UV-1800 spectrometer by Shimadzu JP. The quality of the LAB used for the LS production was confirmed by the RENO experiment and we can expect an attenuation length longer than a few meters, which is good enough for the JSNS$^{2}$ experiment. However, because the LS was stored inside of ISO tanks during transit, additional measurements in Japan were needed to determine if any degradation occurred. The LS was sampled twice in Japan after it was transported from Korea. The first sample was taken immediately after the scintillator arrived in Japan, and the second sample was taken two months later. Both samples were measured at Tohoku University with a 10 cm cuvette. Figure~\ref{fig:transmittance} shows the results of the transmittance measurements made in Korea and Japan. The black solid line is the measurement made in Korea, and the red and blue solid lines are the measurements made in Japan. All three measurements show identical shapes meaning that there was no significant degradation during transportation from Korea to Japan, or during two months of storage inside of an ISO tank. Note that data is normalized to 100 at 440 nm to eliminate the effect of baseline fluctuations of the UV-VIS spectrometers. 

\subsection{Relative Light yield}
\begin{figure}[h]
\begin{tabular}{c c}

\begin{subfigure}{.5\linewidth}
\centering
\includegraphics[scale=0.58]{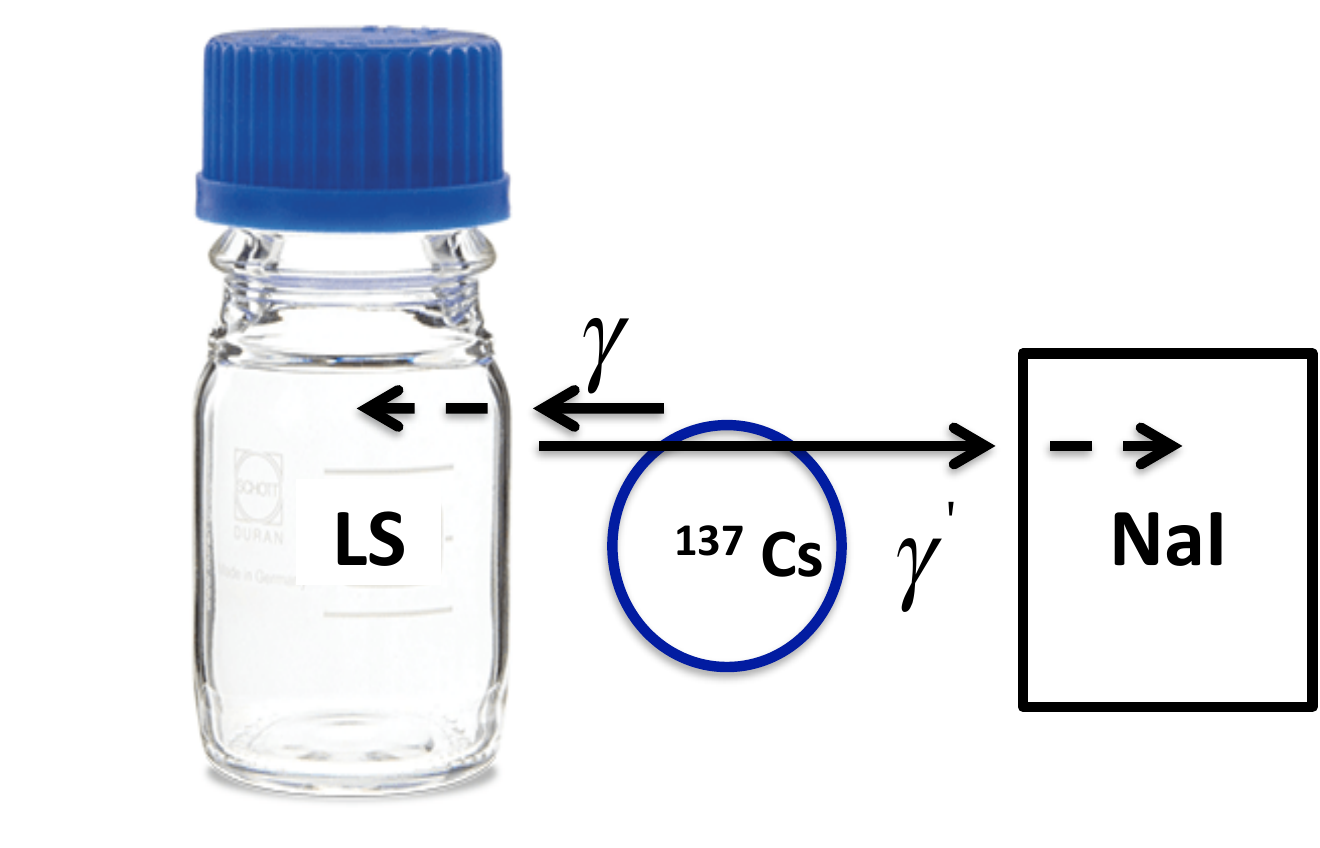}
\caption{The LS light yield measurement setup}
\label{fig:light_yield_a}
\end{subfigure}
&
\begin{subfigure}{.5\linewidth}
\centering
\includegraphics[scale=0.36]{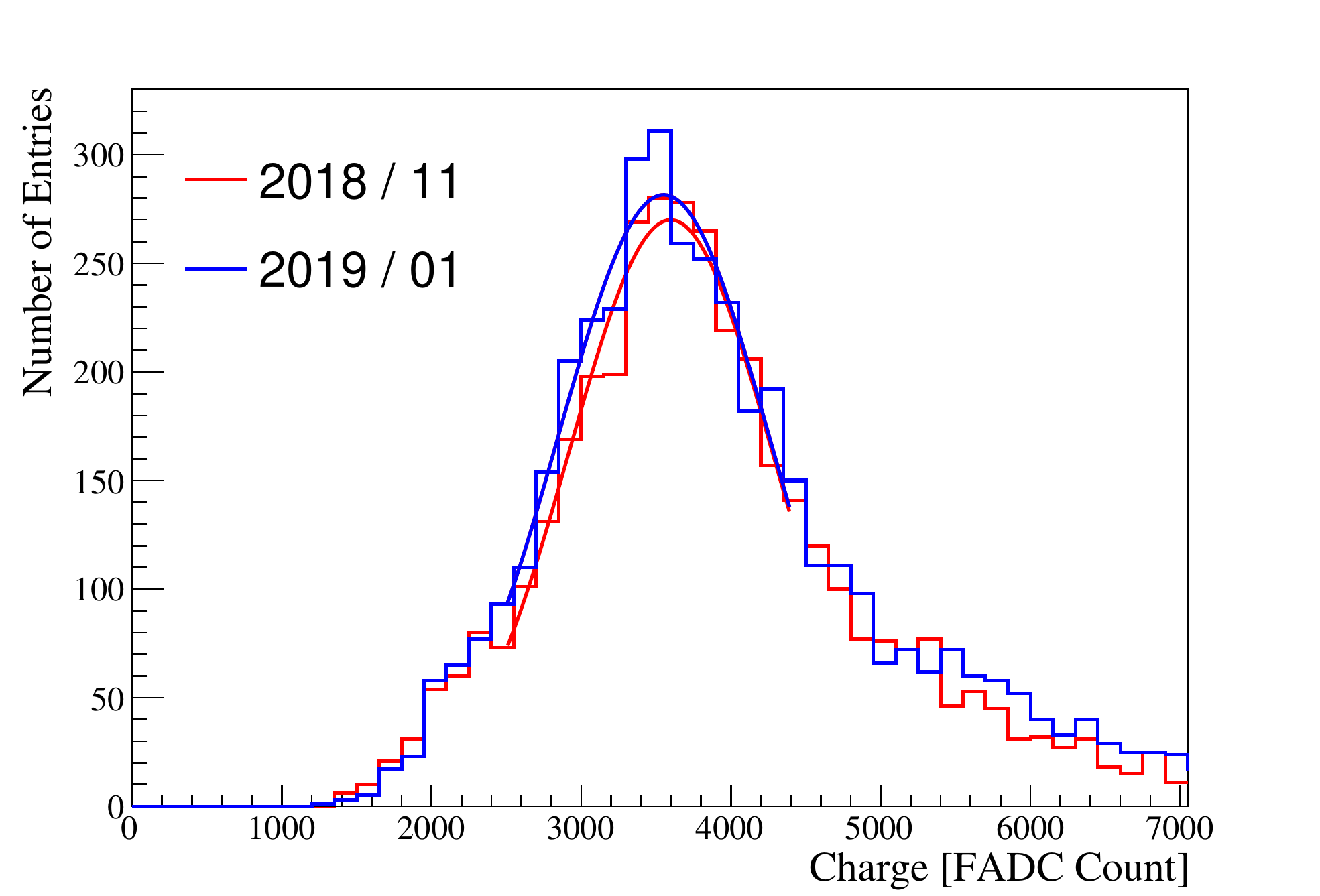}
\caption{A single Gaussian fit after event selection}
\label{fig:light_yield_b}
\end{subfigure}

\end{tabular}

\caption{\setlength{\baselineskip}{4mm} The LS light yield was measured using a coincidence setup between the LS and a NaI crystal. A $^{137}$Cs source was placed between the LS and the NaI crystal. When a gamma emitted from the $^{137}$Cs source enters the LS, it can backscatter and the scattered gamma can enter the NaI crystal. We selected backscattering events from the NaI crystal data and performed a single Gaussian fit to the corresponding LS charge distribution. (a) shows a schematic diagram of the light yield measurement setup, and (b) shows a single Gaussian fit of the LS data with backscattering events selection. The red and blue histograms correspond to different samples of the LS from the same ISO tank taken 2 months apart. The fitted mean value is identical within the statistical uncertainty of fit, meaning there was no significant degradation of the light yield of the LS.} 
\label{fig:light_yield}
\end{figure}

With a small vial ($\sim$100 mL) of LS and a radioactive source, it is difficult to extract an absolute measurement of the light yield because events are not usually completely contained (i.e. no clear energy peak is visible) and there is uncertainty in the optical photon collection efficiency of the photomultiplier tube (PMT). However, the LS production method described above is well-established and our main concern was the possibility of relative degradation due to storage in the ISO tanks. To reduce systematic uncertainty, we used a coincidence setup with LS and a NaI crystal to measure the LS light yield. Figure~\ref{fig:light_yield_a} shows a simplified diagram of the measurement. The measurement was done inside a dark box with two 2-inch PMTs: one attached to the LS and one attached to the NaI crystal. The distance between the LS and the NaI crystal was $\sim$40 cm. We placed a $^{137}$Cs source, which emits monoenergetic 0.662 MeV gamma rays, between the LS and the NaI crystal. When a gamma from $^{137}$Cs enters the LS, it can backscatter and the backscattered gamma can be detected by the NaI crystal. In this case, $\sim$0.48 MeV is deposited into the LS sample and $\sim$0.18 MeV is deposited into the NaI crystal (i.e. the energy loss in the LS and the NaI is Gaussian with peaks at these energies). Waveforms from both the LS and the NaI crystal were collected with a 14 bit Flash Analog-to-Digital Converter with 500 MS/s and a 1 $\mu$s event window. For the LS, the total charge was computed by integrating the waveform from 50 ns before the peak to 250 ns after the peak. For the NaI crystal, the whole waveform was used. After computing the integrated charge, we selected backscattering events from the charge distribution of the NaI crystal. Figure~\ref{fig:light_yield_b} shows the LS charge distribution with a single Gaussian fit after the backscattering event selection. The red and blue histograms are two different samples from the same ISO tank taken 2 months apart. The fitted mean values, 3578 $\pm$ 18 (red) and 3563 $\pm$ 20 (blue, two months later), are the same within statistical error. The similarity in the measured charge distributions indicate that storing the LS in an ISO tank does not result in any significant degradation of the light yield over several months of exposure to the ISO tanks.

\subsection{Temperature monitoring}
\begin{figure}[h]
\begin{center}
\includegraphics[scale=0.5]{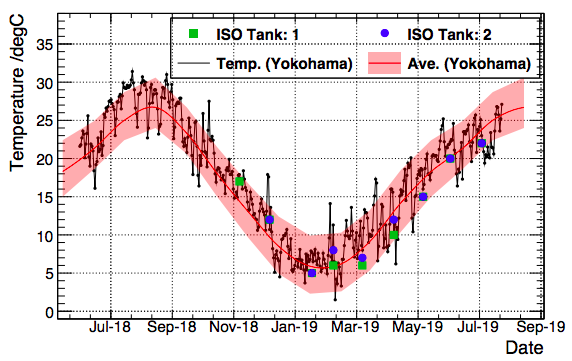}
\end{center}
\caption{\setlength{\baselineskip}{4mm}We measured the temperature of the LS inside each ISO tank every month. The green squares (blue circles) indicate the temperature of the LS inside of ISO tank 1 (2). The red line shows the average temperature in Yokohama, which is the closest meteorological observatory to the ISO tank storage area. The temperature of the LS matches the average temperature in Yokohama quite well, and so we can expect that the temperature of the LS will remain between 0 and 30 degrees. 
} 
\label{fig:LS_temperature}
\end{figure}

The ISO tanks were leased from and stored by a company located in Kawasaki, Japan. A covered storage area was not available, so the ISO tanks stored outside. The ISO tanks are insulated, and therefore it is expected that the temperature of the LS will slowly follow the external temperature. We measured the temperature of the LS inside of each ISO tank every month. Figure~\ref{fig:LS_temperature} shows the temperature variation of the LS. The green squares (blue circles) indicate the temperature of the LS in ISO tank 1 (2). 
We compared the temperature of the LS with the average temperature in Yokohama (red line), which is the closest meteorological observatory with data available from the Japanese government's climate database. These measurements show that the LS temperature is consistent with the average temperature in Yokohama, and we can expect that the LS will not get hotter than $\sim$30 degrees during the summer months. A study of LS aging by JUNO~\cite{cite:JUNO_aging} showed that LS stored at 40 degrees did not show notable degradation. Therefore we do not expect the LS storage in Kawasaki to create any issues for JSNS$^{2}$.

\section{Conclusion}
We produced 38000 L of LS with 3g/L of PPO and 30mg/L of bis-MSB at the RENO LS production site. The site was refurbished and thoroughly cleaned before the production of the LS. The measured water content in the LS was similar to the RENO experiment. Both the relative transmittance and light yield of the LS were measured and showed no significant degradation after transportation from Korea to Japan. The long term stability of the LS properties during storage inside of the ISO tanks was confirmed with measurements of the transmittance and light yield in Japan over a period of several months.

\section*{Acknowledgements}
This work is supported by: the National Research Foundation of Korea (NRF) Grant No. 2009-0083526,  NRF-2017K1A3A7A09016427, and 2018R1D1A1B07050425, funded by the Korea Ministry of Science and ICT. Some of us have been supported by a fund from the BK21 of the NRF; the JSPS grants-in-aid (Grant Number 16H06344, 16H03967), the Ministry of Education, Culture, Sports, Science and Technology (MEXT), Japan.




\end{document}